\documentstyle[12pt,epsfig,graphics,cite,amssymb,amsmath,bm,cool,caption]{article}
\begin{document}\bibliographystyle{plain}\begin{titlepage}
\renewcommand{\thefootnote}{\fnsymbol{footnote}}\hfill
\begin{tabular}{l}HEPHY-PUB 946/14\\UWThPh-2014-34\\December
2014\end{tabular}\\[2cm]\Large\begin{center}{\bf THE SPINLESS
RELATIVISTIC KINK-LIKE PROBLEM}\\[1cm]\large{\bf Wolfgang
LUCHA\footnote[1]{\normalsize\ \emph{E-mail address\/}:
wolfgang.lucha@oeaw.ac.at}}\\[.3cm]\normalsize Institute for High
Energy Physics,\\Austrian Academy of Sciences,\\Nikolsdorfergasse
18, A-1050 Vienna, Austria\\[1cm]\large{\bf Franz
F.~SCH\"OBERL\footnote[2]{\normalsize\ \emph{E-mail address\/}:
franz.schoeberl@univie.ac.at}}\\[.3cm]\normalsize Faculty of
Physics, University of Vienna,\\Boltzmanngasse 5, A-1090 Vienna,
Austria\\[2cm]{\normalsize\bf Abstract}\end{center}\normalsize

\noindent We constrain the possible bound-state solutions of the
spinless Salpeter equation (the most obvious semirelativistic
generalization of the nonrelativistic Schr\"odinger
equation)~with~an interaction between the bound-state constituents
given by the kink-like potential (a central potential of
hyperbolic-tangent form) by formulating a bunch of very elementary
boundary conditions to be satisfied by all solutions of the
eigenvalue problem posed by a bound-state equation of this type,
only to learn that all results produced by a procedure very~much
liked by some quantum-theory practitioners prove to be in
\emph{severe\/} conflict with our expectations.\vspace{6ex}

\vspace{2ex}


\renewcommand{\thefootnote}{\arabic{footnote}}\end{titlepage}

\section{Introduction}\label{Sec:I}Within quantum physics,
nonrelativistic bound states of spinless particles are described
by the time-independent Schr\"odinger equation. A generalization
of such bound-state equation towards a relativistic treatment of
bound states is found if replacing the nonrelativistic free energy
by its relativistically correct counterpart. The outcome of the
improvement is called the spinless Salpeter equation. This name of
the latter derives from the fact that it emerges in the course of
the three-dimensional reduction of the Bethe--Salpeter formalism
\cite{BSE}, which constitutes a commonly accepted formalism for
the Lorentz-covariant description~of~bound states within quantum
field theory: assuming in the homogeneous Bethe--Salpeter equation
all bound-state constituents to propagate freely and interact
instantaneously simplifies this equation to the so-called Salpeter
equation \cite{SE}; disregarding, moreover, the negative-energy
contributions and ignoring the spin degrees of freedom of all
bound-state constituents leads finally to the spinless Salpeter
equation. By construction, such spinless Salpeter equation is the
eigenvalue equation of a Hamiltonian $H$ that combines the sum $T$
of the relativistic free energies of the bound-state constituents
with a potential $V$ that represents the interactions between the
particles forming this bound state. For two particles of masses
$m_1,m_2,$ relative momentum $\bm{p},$ and relative coordinate
$\bm{x},$ this operator $H$ reads (in natural~units $\hbar=c=1$)
\begin{equation}H\equiv T(\bm{p})+V(\bm{x})\ ,\qquad
T(\bm{p})\equiv\sqrt{\bm{p}^2+m_1^2}+\sqrt{\bm{p}^2+m_2^2}\
.\label{Eq:H}\end{equation}

Hamiltonians of the above type are, in general, nonlocal
operators. Hence, finding~exact and, in particular, analytic
solutions to spinless-Salpeter problems is a definitely nontrivial
task. However, one may try to deduce rigorous constraints on the
predicted energy spectra. Here, we do this for what is called the
kink-like potential, a spherically symmetric potential $V(r)$
($r\equiv|\bm{x}|$), depending on a range parameter $\rho$ and a
dimensionless coupling constant~$\kappa$:\begin{equation}
V(\bm{x})=V_{\rm
K}(r)\equiv\kappa\,\rho\tanh(\rho\,r)\equiv\kappa\,\rho\,
\frac{\exp(\rho\,r)-\exp(-\rho\,r)}{\exp(\rho\,r)+\exp(-\rho\,r)}\
,\qquad \rho>0\ ,\qquad \kappa\ge0\ .\label{Eq:VK}\end{equation}
Clearly, this potential function $V_{\rm K}(r)$ is strictly
increasing from $V_{\rm K}(0)=0$ to $V_{\rm
K}(\infty)=\kappa\,\rho$:$$V_{\rm K}(0)=0\le V_{\rm K}(r)\le
V_{\rm K}(\infty)=\kappa\,\rho\ .$$So, this potential is evidently
bounded from below and, therefore, also the Hamiltonian
(\ref{Eq:H}). We get constraints on both energy spectra
(Sec.~\ref{Sec:OI}) and number of bound
states~(Secs.~\ref{Sec:BUB},~\ref{Sec:DUB}).

\section{Eigenvalue Constraints from Operator Inequalities}
\label{Sec:OI}``Very frequently,'' that is, in fact, in almost all
instances, it proves impossible to determine, by \emph{analytic\/}
means, energy levels of given bound-state problems in quantum
physics \emph{exactly}. In such situation, the location of the
energy eigenvalues may be constrained by comparison with
bound-state problems for which knowledge about their eigenvalue
spectra is available. A valuable tool in any endeavour of this
kind is the following \emph{spectral comparison
theorem\/}:\footnote{For a proof, see
Refs.~\cite[Sec.~III]{Lucha96:RCP}, \cite[Sec.~3]{Lucha99:talk1},
\cite[Sec.~2]{Lucha99:talk2}, \cite[Appendix]{Lucha99},
\cite[Subsec.~3.1]{Lucha04:TWR}, or \cite[Appendix]{RHO05}.} For
any pair of self-adjoint semibounded operators $A$ (with
eigenvalues $a_k,$ $k\in{\mathbb N}_0,$ ordered by $a_0\le a_1\le
a_2\le\cdots$) and $B$ (with eigenvalues $b_k,$ $k\in{\mathbb
N}_0,$ ordered by $b_0\le b_1\le b_2\le\cdots$) satisfying the
inequality $A\le B,$ the eigenvalues below onset of the essential
spectrum~fulfil$$a_k\le b_k\ ,\qquad k\in{\mathbb N}_0\ .$$We find
numerous possibilities for applying this theorem to a
semirelativistic Hamiltonian, in general, or to the particular
instance of bound-state problems with a kinky
potential~(\ref{Eq:VK}).

By definition, the relativistic free-energy term $T(\bm{p})$
defined in Eq.~(\ref{Eq:H}) and --- because of $V_{\rm K}(r)\ge0$
--- the Hamiltonian $H$ with interaction potential (\ref{Eq:VK})
satisfy trivial lower~bounds:$$T(\bm{p})\ge m_1+m_2\ge0\ ,\qquad
H\ge m_1+m_2\ge0\ .$$Therefore, all eigenvalues $E_k,$
$k=0,1,2,\dots,$ of our operator $H$ are bounded from
below~by$$E_k\ge m_1+m_2\ge0\ ,\qquad k=0,1,2,\dots\ ;$$hence, its
\emph{binding\/} energies $B_k\equiv E_k-m_1-m_2,$
$k=0,1,2,\dots,$ are necessarily \emph{non-negative}:$$B_k\ge0\
,\qquad k=0,1,2,\dots\ .$$

A trivial upper bound to each semirelativistic Hamiltonian
(\ref{Eq:H}) arises from the fact that, regarded as functions of
$\bm{p}^2,$ the Schr\"odinger free term $T_{\rm NR}(\bm{p})$ found
as nonrelativistic (NR) limit is linear and tangent to the sum of
square roots in the relativistic kinetic~energy~$T(\bm{p})$:
$$T(\bm{p})\le T_{\rm NR}(\bm{p})\equiv
m_1+m_2+\frac{\bm{p}^2}{2\,m_1}+\frac{\bm{p}^2}{2\,m_2}\qquad
\Longrightarrow\qquad H\le H_{\rm NR}\equiv T_{\rm
NR}(\bm{p})+V(\bm{x})\ .$$Our spectral comparison theorem then
tells us that any Schr\"odinger energy eigenvalue $E_{{\rm S},k}$
is an upper bound to its spinless-Salpeter counterpart $E_k$:
$E_k\le E_{{\rm S},k},$ $k=0,1,2,\dots.$ Thus, for a given
potential, there is a bound state of $H$ below each bound state of
$H_{\rm NR}.$ Hence,~the number of bound states of $H,$ $N,$ is
not lower than the number $N_{\rm NR}$ of bound~states of~$H_{\rm
NR}$:$$N\ge N_{\rm NR}\ .$$

For our quest, it is convenient and advantageous to define the
shifted kink-like potential\begin{equation}\widetilde V_{\rm
K}(r)\equiv V_{\rm K}(r)-\kappa\,\rho=\kappa\,\rho
\left[\tanh(\rho\,r)-1\right]=-\frac{2\,\kappa\,\rho}
{1+\exp(2\,\rho\,r)}\ ,\label{Eq:VKt}\end{equation}which is
negative and monotonically increasing for $r<\infty,$ and
approaches zero for $r\to\infty$:$$\widetilde V_{\rm
K}(0)=-\kappa\,\rho\le \widetilde V_{\rm K}(r)\le\widetilde V_{\rm
K}(\infty)=0\ ,\qquad\lim_{r\to\infty}\widetilde V_{\rm K}(r)=0\
.$$Such shift has no implications on basic characteristics of
bound-state problems, such as the number of bound states, and
evident ones, easily taken into account, on energy~eigenvalues.
Finding simple potentials that form either an upper or a lower
bound to the (shifted~or~not) kink-like potential does not pose a
particularly big challenge. We discuss but a few of
these:\begin{itemize}\item As border case, a candidate suggesting
itself for comparison is the Coulomb potential$$V_{\rm
C}(r)\equiv-\frac{\alpha}{r}\ ,\qquad0\le\alpha<\alpha_{\rm
c}=\frac{4}{\pi}=1.273239\dots\ .$$Its critical coupling
$\alpha_{\rm c}$ arises from demanding the relativistic Coulomb
problem~to~be bounded from below \cite{Herbst}. This potential is
obviously a lower bound to $\widetilde V_{\rm K}(r)$ if
$\alpha=\kappa$:$$V_{\rm C}(r)\le \widetilde V_{\rm
K}(r)\qquad\mbox{for}\ \alpha=\kappa\ .$$This lower bound can be
optimized by diminishing the value of the Coulomb coupling from
$\alpha=\kappa$ until Coulomb potential and shifted kinky
potential $\widetilde V_{\rm K}(r)$ get in contact. Requesting, at
the point of contact, equality of the potentials and of their
derivatives, we obtain a class of Coulomb lower bounds to
$\widetilde V_{\rm K}(r)$ for all Coulomb couplings
$\alpha\ge\overline{\alpha},$ where the touching Coulomb coupling
$\overline{\alpha}$ is the solution of the
$\rho$-independent~equation$$1+\frac{\overline{\alpha}}{\kappa}
=\log\frac{\kappa}{\overline{\alpha}}\qquad\Longrightarrow\qquad
\overline{\alpha}\le\kappa\ .$$For the relativistic Coulomb
problem, in turn, rigorous lower bounds to the spectrum
$\sigma(H)$ of the operator $H$ could be given
\cite{Herbst,Martin89} for the equal-mass case $m_1=m_2=m$:
\begin{align*}\sigma(H)&\ge2\,m\,
\sqrt{1-\left(\frac{\alpha}{\alpha_{\rm c}}\right)^{\!2}}=2\,m\,
\sqrt{1-\left(\frac{\pi\,\alpha}{4}\right)^{\!2}}&\quad&\mbox{for}\
\alpha<\alpha_{\rm c}\ ,\\\sigma(H)&\ge2\,m\,
\sqrt{\frac{1+\sqrt{1-\alpha^2}}{2}}&\quad&\mbox{for}\ \alpha\le1\
.\end{align*}\item An exponential potential with the same
potential parameters as in the kink-like~case,$$V_{\rm E}(r)
\equiv-\kappa\,\rho\exp(-\rho\,r)\ ,$$constitutes a lower bound to
the kink-like potential (\ref{Eq:VK}), as is straightforward
to~show:$$V_{\rm E}(r)\le\widetilde V_{\rm K}(r)\ .$$\item
Likewise, it is an easy task to convince oneself that the
exponential-squared potential$$V_{\rm E2}(r)\equiv-\kappa\,\rho
\exp(-2\,\rho\,r)\ ,$$\emph{i.e.}, with slope twice as large as in
Eq.~(\ref{Eq:VK}), is an upper limit to the kink-like potential:
$$\widetilde V_{\rm K}(r)\le V_{\rm E2}(r)\ .$$\end{itemize}

\section{Number of Bound States of Schr\"odinger Problems}
\label{Sec:BUB}Estimating the maximum number $N_{\rm NR}$ of bound
states accommodated by a nonrelativistic bound-state problem is,
in view of the pertinent results available, not exorbitantly
difficult. Among the first results in this respect is the bound by
Bargmann \cite{Bargmann}: For the Hamiltonian
$$H=\frac{\bm{p}^2}{2\,\mu}+V(r)\ ,\qquad\mu>0\ ,$$with $\mu$
given either by $\mu=m,$ for a single bound particle of mass $m,$
or by the reduced~mass$$\mu=\frac{m_1\,m_2}{m_1+m_2}$$of a system
of two bound particles of masses $m_1,m_2,$ Bargmann finds, as
constraint to~$N_{\rm NR},$ $$N_{\rm NR}\lneqq\frac{I\,(I+1)}{2}\
,\qquad I\equiv2\,\mu\int_0^\infty{\rm d}r\,r\,|V_-(r)|\ ,$$where
just all negative potential parts matter:
$V_-(r)\equiv-\max[0,-V(r)]=V(r)\,\theta(-V(r)).$ For a shifted
kinky potential (\ref{Eq:VKt}), $I$ has to be proportional to
$\mu\,\kappa/\rho$ on dimensional grounds:
$$I=2\,\mu\int_0^\infty{\rm d}r\,r\,|\widetilde V_{\rm K}(r)|
=\frac{\pi^2\,\mu\,\kappa}{12\,\rho}\ ,\qquad N_{\rm NR}
\lneqq\frac{\pi^2\,\mu\,\kappa}{24\,\rho}
\left(\frac{\pi^2\,\mu\,\kappa}{12\,\rho}+1\right).$$

\section{Maximal Number of Semirelativistic Bound States}
\label{Sec:DUB}For a class of bound-state equations including the
spinless Salpeter equation, I.~Daubechies has found an
easy-to-apply upper limit to the total number $N$ of bound states
\cite{Daubechies}. Assume that the Hamiltonian controlling the
system under consideration, $H\equiv K(|\bm{p}|)+V(\bm{x}),$ is an
$L^2({\mathbb R}^3)$ operator composed of two sufficiently
restricted ingredients: a kinetic energy~$K(|\bm{p}|)$ that is a
strictly increasing, differentiable function of only the modulus
of $\bm{p}$ and a potential $V(\bm{x})$ that is a smooth function
of compact support, $V\in C^\infty_0({\mathbb R}^3),$ satisfying
the conditions$$K(|\bm{p}|)\ge0\ ,\qquad K(0)=0\ ,\qquad
\lim_{|\bm{p}|\to\infty}\!\!K(|\bm{p}|)=\infty\ ,\qquad
V(\bm{x})\le0\ .$$Under these conditions, the total number $N$ of
bound states of $H$ is bounded from above~by
\begin{equation}N\le\frac{C}{6\,\pi^2}\int{\rm d}^3x
\left[K^{-1}(|V(\bm{x})|)\right]^3\ ;\label{Eq:DB}\end{equation}
the constant $C,$ converting semiclassical into quantum limit
\cite{Daubechies}, can be found numerically:$$C=\inf_{b>0}\!
\left( \left\{e^b\int_0^\infty\frac{{\rm d}y}{y^2}\,e^{-b\,y}
\,[g(y)]^3\right\} \left\{b\int_0^\infty\frac{{\rm d}y\,y}{y+1}\,
e^{-b\,y}\right\}^{\!-1}\right),\qquad g(y)\equiv\sup_{x>0}
\frac{K^{-1}(x\,y)}{K^{-1}(x)}\ .$$For two particles of arbitrary
masses $m_1,m_2,$ the \emph{relativistic kinetic\/} term,
$K(|\bm{p}|),$ becomes\footnote{For relativistic kinetic terms,
the bound (\ref{Eq:DB}) holds for any potential $V(\bm{x})\le0$ in
$L^{3/2}({\mathbb R}^3)\cap L^3({\mathbb
R}^3)$~\cite{Daubechies}.}$$K(|\bm{p}|)
=\sqrt{|\bm{p}|^2+m_1^2}+\sqrt{|\bm{p}|^2+m_2^2}-m_1-m_2\ .$$The
inverse of this free-energy function, $K^{-1}(x),$ required by the
bound (\ref{Eq:DB}), is easily~found:$$K^{-1}(x)=
\frac{\sqrt{x\,(x+2\,m_1)\,(x+2\,m_2)\,(x+2\,m_1+2\,m_2)}}
{2\,(x+m_1+m_2)}\ ,\qquad x\ge0\ .$$For the special case of
\emph{equal\/} masses, \emph{i.e.}, if $m_1=m_2=m,$ this inverse
function simplifies~to$$K^{-1}(x)=\frac{\sqrt{x\,(x+4\,m)}}{2}\
,\qquad x\ge0\ .$$So, the total number of bound states of the
two-particle spinless-Salpeter equation satisfies$$N\le
\frac{C}{12\,\pi}\int_0^\infty{\rm d}r\,r^2
\left[|V(r)|\left(|V(r)|+4\,m\right)\right]^{3/2}\ ,$$with the
conversion factor $C=14.107590867$ for $m>0$ or $C=6.074898097$
for $m=0$ \cite{Lucha14:SRWSP}.

\section{Application to Approximate Bound-State Solution}The
kink-like potential (\ref{Eq:VK}) has met interest in the context
of the Dirac equation \cite{Grosse92,Castro06},~the Klein--Gordon
equation \cite{Castro07}, and also the spinless Salpeter equation
\cite{Hassanabadi_CPC}. So, let us examine whether the outcomes of
Ref.~\cite{Hassanabadi_CPC} fit to the general restrictions
collected in Secs.~\ref{Sec:OI}, \ref{Sec:BUB},~and~\ref{Sec:DUB}.

In order to define a given spinless Salpeter problem under
consideration unambiguously and completely, the numerical values
of the relevant mass and potential parameters have to be
specified. Basically for illustrative purposes and simplicity of
notation, below the case of bound-state constituents of equal
masses, \emph{i.e.}, $m_1=m_2,$ will be in the focus of our
interest. In Ref.~\cite{Hassanabadi_CPC}, the numerical results
for the binding energies are presented in form of one table and
three figures, each of these relying on a (partly) different
choice of the numerical~values of the involved parameters required
as input; we reproduce these parameter sets in
Table~\ref{Tab:HYP}. To facilitate comparison, we consider
parameter values compatible with the sets in Table~\ref{Tab:HYP}.

\begin{table}[h]\caption{Numerical values adopted in
Ref.~\cite[Table~1 and Figs.~1--3]{Hassanabadi_CPC} for the masses
$m_1,$ $m_2$ of the bound-state constituents and the parameters
$\rho$ and $\kappa$ of the ``kink-like''~potential~(\ref{Eq:VK}).}
\begin{center}\begin{tabular}{lllll}\hline\hline\\[-1.5ex]
\multicolumn{1}{c}{Set of results}&\multicolumn{1}{c}{$m_1$
(arb.~unit)}&\multicolumn{1}{c}{$m_2$ (arb.~unit)}&
\multicolumn{1}{c}{$\rho$ (arb.~unit)}&
\multicolumn{1}{c}{$\kappa$}\\[1.3ex]\hline\\[-1.5ex]
Table~1&2.0&0.5&0.01&0.01\\Figure~1&0.5&0.5&0.0001&0--0.1\\
Figure~2&0--2.0&0.5&0.0001&0.1\\Figure~3&2.0&0--2.0&0.0001&0.1
\\[1.3ex]\hline\hline\end{tabular}\end{center}\label{Tab:HYP}
\end{table}

Having succeeded to formulate a well-defined bound-state problem,
one of the very first questions that arise is that of the mere
number of bound states to expect as solutions of this problem. To
give but an idea, we list in Table \ref{Tab:NBS} the maximal
numbers of bound states~of the relativistic kink-like problem for
specific parameter values within their intervals
in~Table~\ref{Tab:HYP}: we obtain fairly large numbers of possible
bound states, primarily owing to the fact~that~for the parameter
values adopted in Ref.~\cite{Hassanabadi_CPC}, particularly that
of the range $\rho$ in the figures, the kink-like potential is
extremely shallow. All binding energies can vary only over
the~interval$$V_{\rm K}(0)=0\le B_k\le V_{\rm K}(\infty)
=\kappa\,\rho\ ,\qquad k=0,1,2,\dots\ .$$For $\rho=0.0001$ and
$\kappa=0.1,$ this is really tiny, compared to masses of $O(1)$:
$0\le B_k\le10^{-5}.$

\begin{table}[h]\caption{Upper bounds $\overline{N}\ge N$
($\overline{N}_{\rm NR}\ge N_{\rm NR}$), computed along the lines
discussed in Sec.~\ref{Sec:DUB} (Sec.~\ref{Sec:BUB}), to the
number $N$ ($N_{\rm NR}$) of bound states caused by the
relativistic (nonrelativistic) Hamiltonian $H$ ($H_{\rm NR}$)
defined in Sec.~\ref{Sec:I} (Sec.~\ref{Sec:OI}) with the kink-like
potential (\ref{Eq:VK}), for selected numerical values of mass and
potential parameters used by Ref.~\cite[Table~1 and
Figs.~1--3]{Hassanabadi_CPC}.}
\begin{center}\begin{tabular}{lllllrr}\hline\hline\\[-1.5ex]
\multicolumn{1}{c}{Set of results}&\multicolumn{1}{c}{$m_1$
(arb.~unit)}&\multicolumn{1}{c}{$m_2$ (arb.~unit)}&
\multicolumn{1}{c}{$\rho$ (arb.~unit)}&
\multicolumn{1}{c}{$\kappa$}&\multicolumn{1}{c}{$\overline{N}$}&
\multicolumn{1}{c}{$\overline{N}_{\rm NR}$}\\[1.3ex]\hline
\\[-1.5ex]Table~1&2.0&0.5&0.01&0.01&\multicolumn{1}{c}{---}&0\\
Figure~1&0.5&0.5&0.0001&0.01&171&221\\
Figure~2&0.5&0.5&0.0001&0.1&5419&21241\\
Figure~3&2.0&2.0&0.0001&0.1&43356&338637\\[1.3ex]\hline\hline
\end{tabular}\end{center}\label{Tab:NBS}\end{table}

Our logically next move must be to narrow down the location of the
energy levels~of the problem under study with a sufficient degree
of rigour. This can be achieved, among others, by exploitation of
variational techniques or envelope theory, reviewed in,
\emph{e.g.},~Refs.~\cite{Lucha01-DMAIa,Lucha01-DMAIb,Lucha02-DMAII};
for recent use of Rayleigh--Ritz methods in spinless-Salpeter
problems, see
Refs.~\cite{Lucha14:SRWSP,Lucha14:Q@W,Lucha14:SRHP,Lucha14:SRYP}.

The (asserted) ultimate outcomes of the sequence of simplifying
assumptions applied in Ref.~\cite{Hassanabadi_CPC} in order to be
able to find a kind of approximate solution to the spinless
relativistic kink-like problem, for bound states of vanishing
orbital angular momentum~only, consist~of\begin{itemize}\item a
quadratic relation, Eq.~(26) of Ref.~\cite{Hassanabadi_CPC}, for
the bound states' binding energies,~with the explicit expressions
for its pair of solutions given in Eq.~(27) of
Ref.~\cite{Hassanabadi_CPC}, as well~as\item an associated set of
(unnormalized) bound-state wave functions, Eq.~(37) of
Ref.~\cite{Hassanabadi_CPC}.\end{itemize}

However, the \emph{supposedly\/} exact solutions to the spinless
Salpeter equation with kink-like potential proposed in
Ref.~\cite{Hassanabadi_CPC} give rise to considerable concerns.
Although --- as recalled in Sec.~\ref{Sec:OI} --- all binding
energies emerging from a spinless Salpeter equation with
non-negative potential (which holds for the kink-like potential,
\emph{cf}.~Sec.~\ref{Sec:I}) are, already by definition of the
potential, non-negative, strangely enough the numerical values of
all binding energies given in Ref.~\cite[Table~1 and
Figs.~1--3]{Hassanabadi_CPC} are (upon reinstalling obviously
missing negative~signs in the labels of the ordinate of Fig.~3 in
Ref.~\cite{Hassanabadi_CPC}) negative. Consequently, among the
results of Ref.~\cite{Hassanabadi_CPC} there is nothing left for
us to compare with, since acceptable binding energies~have to be
strictly positive. As a matter of fact, already the most cursory
inspection reveals~that, irrespective of parameters used,
\emph{both\/} roots of Eq.~(26) of Ref.~\cite{Hassanabadi_CPC} are
bound to be negative.

\section{Summary and Concluding Remarks}Refraining from dealing
with just numerical or analytical but just approximate solutions
to the spinless Salpeter equation for which the achieved accuracy
is not entirely under control, we formulated for the
semirelativistic bound-state problem posed by such spinless
Salpeter equation with hyperbolic-tangent-shaped ``kink-like''
central potential a couple of rigorous, but still elementary
boundary conditions each corresponding exact solution should
respect. Astonishingly, a recent study of precisely this problem
\cite{Hassanabadi_CPC}, following a rather popular (and thus in
these surroundings frequently adopted) route of approximations
comes forth with a tentative solution that fails in satisfying
already the most trivial of our general~constraints.

\end{document}